\documentclass[conference]{IEEEtran}
\IEEEoverridecommandlockouts
\usepackage{cite}
\usepackage{amsmath,amssymb,amsfonts}
\usepackage{algorithmic}
\usepackage{graphicx}
\usepackage{epstopdf}
\DeclareGraphicsExtensions{.png,.pdf}
\usepackage{textcomp}
\usepackage{xcolor}
\usepackage{url}
\usepackage{listings}
\def\BibTeX{{\rm B\kern-.05em{\sc i\kern-.025em b}\kern-.08em
    T\kern-.1667em\lower.7ex\hbox{E}\kern-.125emX}}
\begin{document}

\title{Exosphere - Bringing The Cloud Closer\\
}

%\author{\IEEEauthorblockN{Julian Pistorius}
%\IEEEauthorblockA{\textit{CALS Comm and Technologies} \\
%\textit{The University of Arizona}\\
%Tucson, Arizona, USA \\
%ORCID: 0000-0002-3485-0084}
%\and
%\IEEEauthorblockN{Chris Martin}
%\IEEEauthorblockA{\textit{Founding Developer and Maintainer} \\
%\textit{Exosphere Project}\\
%Flagstaff, Arizona, USA \\
%chris@c-mart.in}
%\and
%\IEEEauthorblockN{Sanjana Sudarshan}
%\IEEEauthorblockA{\textit{Research Technologies} \\
%\textit{Indiana University}\\
%Bloomington, Indiana, USA \\
%ssudarsh@iu.edu}
%}
%\and
%\IEEEauthorblockN{David S. LeBauer}
%\IEEEauthorblockA{\textit{College of Agriculture and Life Sciences} \\
%\textit{The University of Arizona}\\
%Tucson, Arizona, USA \\
%ORCID: 0000-0001-7228-053X\\
%dlebauer@arizona.edu}

\author{
\IEEEauthorblockN{
Julian Pistorius \IEEEauthorrefmark{1} \IEEEauthorrefmark{4}, 
Chris Martin\IEEEauthorrefmark{2}, 
Sanjana Sudarshan\IEEEauthorrefmark{3} and 
David S. LeBauer\IEEEauthorrefmark{1}\IEEEauthorrefmark{6}}
\IEEEauthorblockA{
\IEEEauthorrefmark{1}College of Agriculture and Life Sciences -- The University of Arizona -- Tucson, Arizona, USA\\
\IEEEauthorrefmark{4}julianp@arizona.edu, ORCID:0000-0002-3485-0084\\
\IEEEauthorrefmark{6}dlebauer@arizona.edu, ORCID:0000-0001-7228-053X}
\IEEEauthorblockA{
\IEEEauthorrefmark{2}Founding Developer and Maintainer -- Exosphere Project -- Flagstaff, Arizona, USA, chris@c-mart.in}
\IEEEauthorblockA{
\IEEEauthorrefmark{3}Research Technologies -- Indiana University -- Bloomington, Indiana, USA, ssudarsh@iu.edu}}
%}

\maketitle

\begin{abstract}
Exosphere provides researcher-friendly software for managing computing workloads on OpenStack cloud infrastructure. Exosphere is a user-friendly alternative to Horizon, the default OpenStack graphical interface. Exosphere can be used with most research cloud infrastructure, requiring near-zero custom integration work.
\end{abstract}

\begin{IEEEkeywords}
Cloud computing, Extreme Science and Engineering Discovery Environment (XSEDE), Jetstream Cloud, OpenStack, user interface (UI), usability
\end{IEEEkeywords}

\section{Background and Motivation}

Researchers use cloud services for on-demand and interactive scientific computing workloads. Many institutions are establishing cloud infrastructure and others are increasing the capacity of their existing clouds. OpenStack \cite{noauthor_openstack_nodate} is the operating system which powers many of these research clouds, including all of the following:

\begin{itemize}
\item Recent NSF award recipients: Jetstream 2, Massachusetts Open Cloud, CloudLab, Chameleon Cloud, and Aristotle Cloud Federation
\item Campus clouds at many research institutions (commissioning now: SUNY Binghamton and Buffalo, University of Cincinnati, UC Santa Barbara)
\item Research clouds at Oak Ridge and Los Alamos National Laboratories
\item International research organizations (CERN, New Zealand eScience Infrastructure, Australia's National eResearch Collaboration Tools and Resources project (Nectar))
\end{itemize}

Research cloud services must provide user-friendly interfaces in order to broaden adoption within scientific communities, particularly by researchers who do not have a strong computational background. Unfortunately, the default user interface for OpenStack (named Horizon) was developed for use by IT system administrators. To complete common tasks (e.g. creating and connecting to a cloud instance), Horizon demands familiarity with firewall security groups, SSH key pairs, and computer networking. Atmosphere (developed by CyVerse)\cite{skidmore_iplant_2011} is a user-friendly alternative interface to OpenStack. Atmosphere’s source code is freely available, but each Atmosphere deployment requires significant integration work and approximately one full-time developer in ongoing maintenance and specialized user support. Most groups that host research clouds cannot support Atmosphere, so only two organizations (CyVerse and Jetstream\cite{stewart_jetstream_2015, towns_xsede_2014}) use it in production. With Horizon left as the only widely-deployed graphical interface for research clouds, there is a gap between the potential of these resources and scientists' ability to use them.

\section{Proposed Solution}

We address this problem with Exosphere: a researcher-friendly software client for OpenStack which lowers the barrier to using powerful and flexible cloud services. Exosphere is designed to be used with any OpenStack cloud without custom integration work or maintenance by that cloud's administrator. Exosphere communicates with OpenStack services: the cloud administrator usually does not need to do anything except provide access to the standard OpenStack APIs. Community members already use Exosphere to access Jetstream Cloud and CyVerse's OpenStack infrastructure. Exosphere is fully open-source (BSD-licensed) software.

Exosphere can support research computing applications in many ways, empowering users to create and manage persistent servers for databases and science gateways, scalable clusters for scientific computing, classrooms and workshops, and disposable sandbox servers for maximum flexibility in testing and development work.

\subsection{Goals and values}

\begin{itemize}
    \item Deliver the most user-friendly way to manage workloads on non-proprietary cloud infrastructure
    \item Provide a consistent user experience across infrastructures operated by different organizations
    \item Empower people to use Exosphere along with other tools to manage the same resources, with minimal cost of switching (i.e. make it easy to move away from Exosphere to other tools, and come back at will)
    \item Support security-focused workloads
    \item Promote an open, community-driven development approach (e.g. user interviews\cite{noauthor_user_nodate} generate issues tagged with "Community request"\cite{noauthor_issues_nodate})
\end{itemize}

\section{Architecture}

\begin{figure*}[t!]
\centerline{\includegraphics[width=1.1\textwidth]{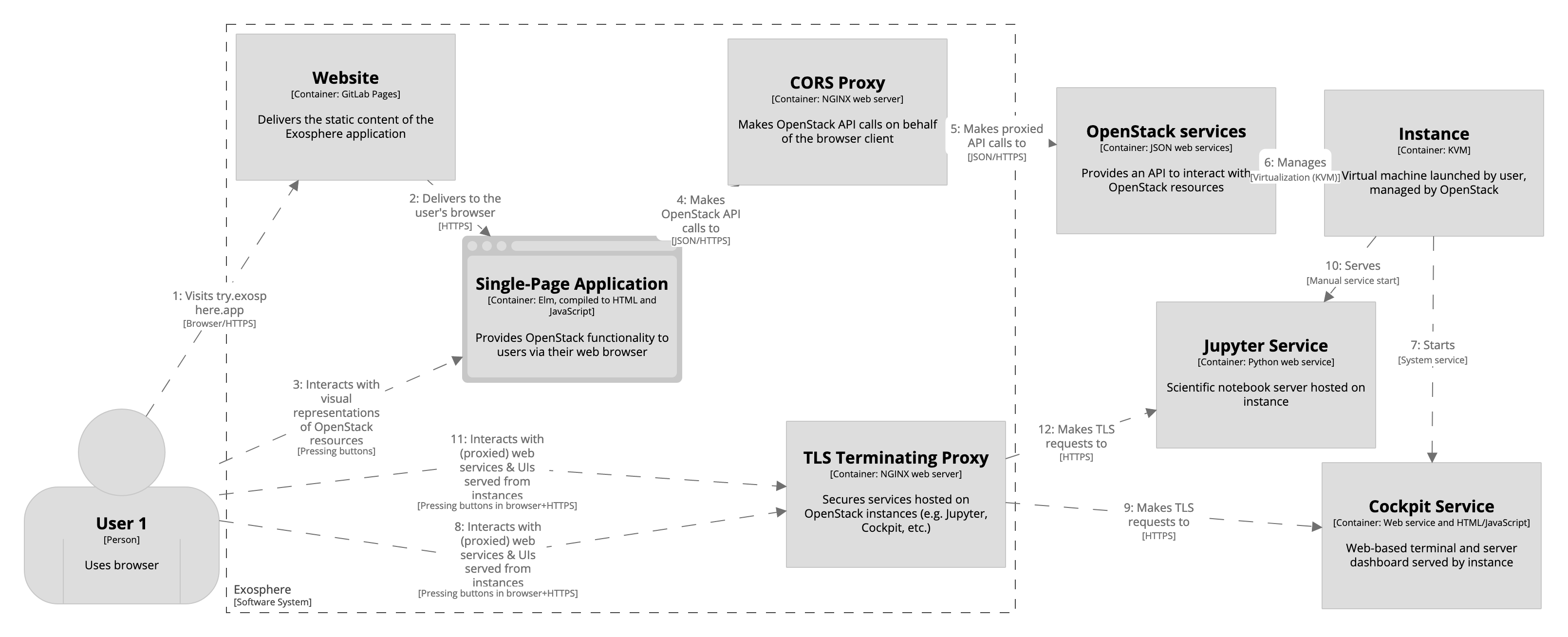}}
\caption{Exosphere Architecture Diagram.}
\label{fig:exo-architecture}
\end{figure*}

The Exosphere client application is usable against OpenStack with no other dependencies on back-end services, although proxy servers are used where needed (Fig. \ref{fig:exo-architecture}). These proxies facilitate secure connections to services running on users' instances, and connections to OpenStack APIs when they would otherwise be unreachable due to restrictions imposed by the user's web browser (same-origin policy) or the user's network (blocking outbound TCP ports used by OpenStack APIs).

Exosphere is written with Elm\cite{czaplicki_elm_nodate}, a functional programming language for building user interfaces. It is compiled to JavaScript, HTML, and CSS. The same codebase is usable in a standard web browser, or as a cross-platform desktop application (currently using the Electron\cite{noauthor_electron_nodate} framework). 

\section{Features}

% cmart response to Jeremy feedback: I think these features are relatively self-explanatory and we don't need prose to describe what the intended audience can probably understand with a few bullets. I think turning these into prose would add verbosity with little benefit.

\subsection{Features available now}

\begin{figure}[bhtp]
\centerline{\includegraphics[width=0.5\textwidth]{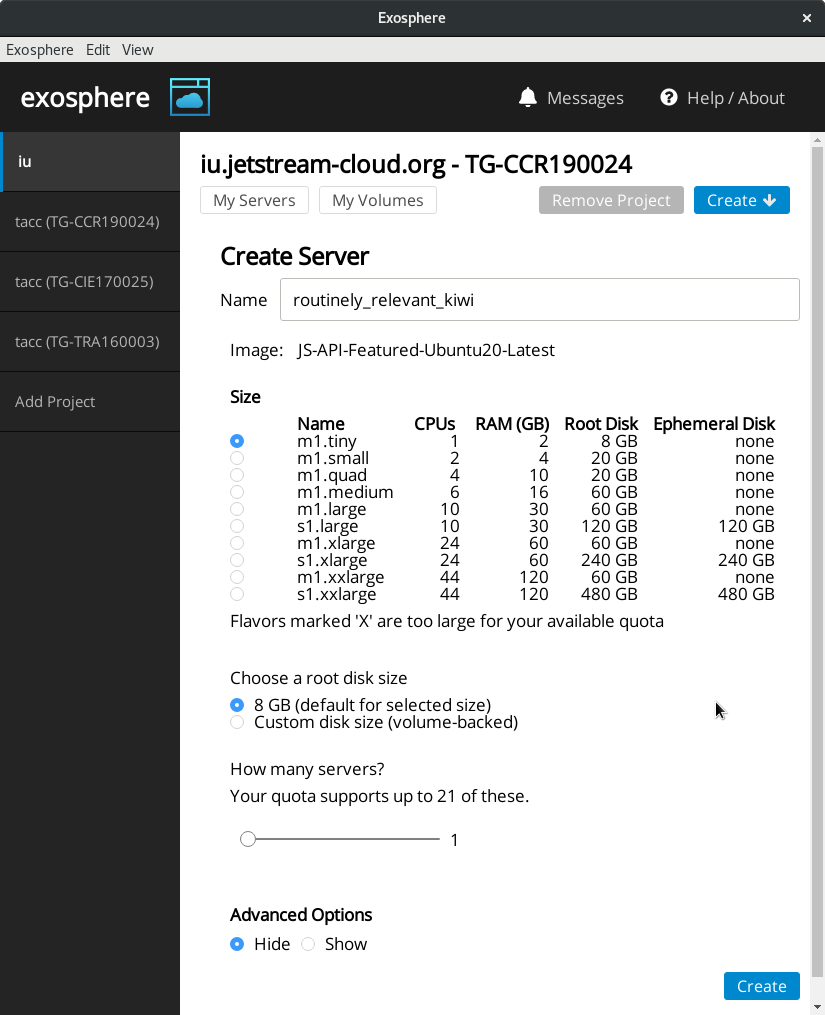}}
\caption{Launching a virtual server in Exosphere.}
\label{figlaunch}
\end{figure}

\begin{figure*}[ptbh]
\centerline{\includegraphics[width=0.95\textwidth]{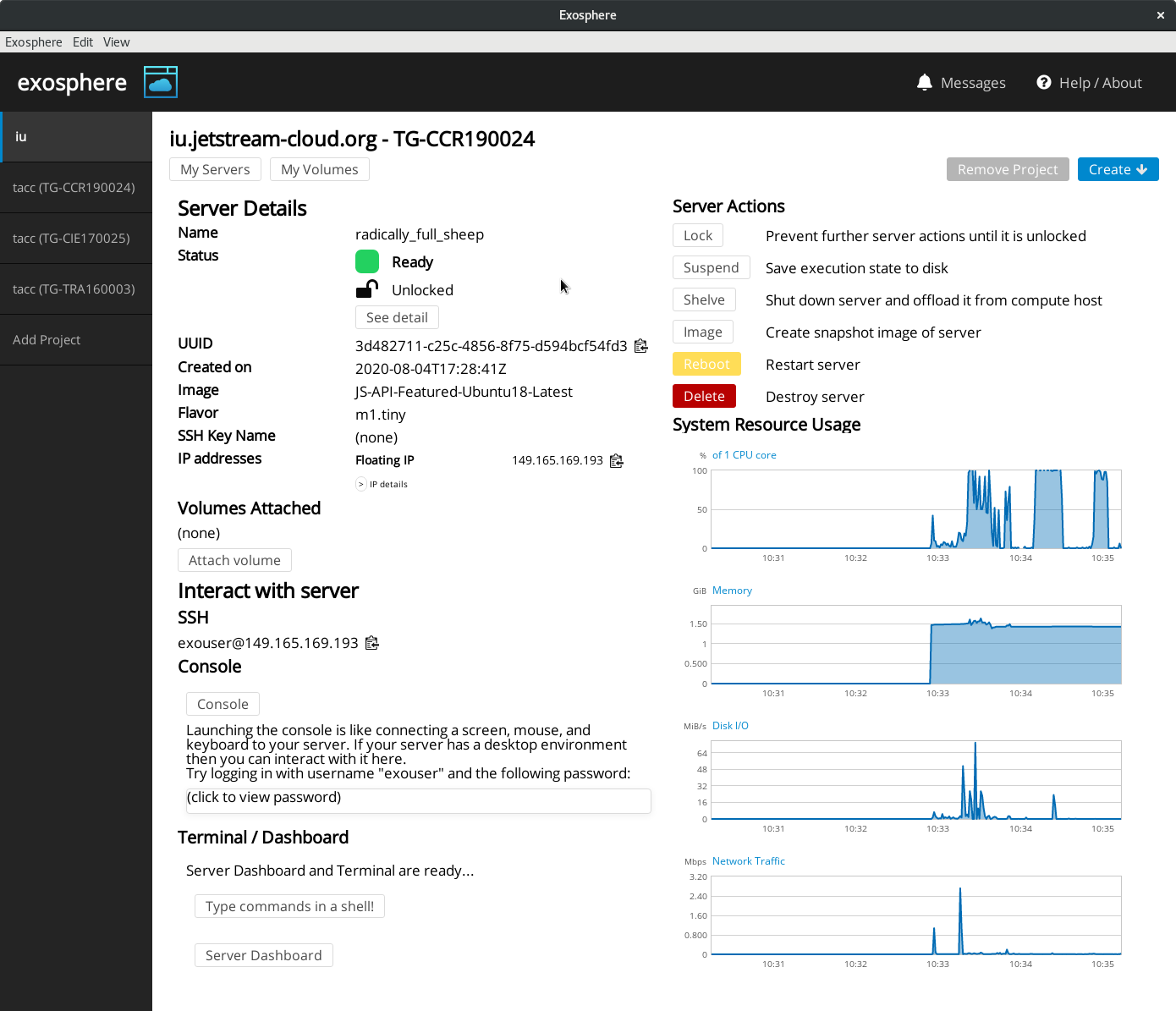}}
\caption{Exosphere's Server Details view.}
\label{figdetails}
\end{figure*}

\begin{itemize}
    \item Create and manage cloud instances and volumes (Fig.~\ref{figlaunch}, Fig.~\ref{figdetails}).
    \item Delivers on each instance, using Cockpit\cite{noauthor_cockpit_nodate}:
    \begin{itemize}
        \item One-click terminal, no knowledge of SSH required
        \item One-click graphical dashboard
    \end{itemize}
    \item Usable with nearly any OpenStack cloud\footnote{Queens release or newer, with service APIs accessible via network from the user's device}.
    \item Completely standalone app, no custom back-end server required for use of core features.
    \item Secure defaults:
        \begin{itemize}
        \item No administrative backdoor SSH key needs to be deployed on instances.
        \item No need for application to have administrator access to OpenStack; works with standard user privileges.
        \item Small, well-defined set of dependencies and "moving parts".
        \item Continuous Integration (CI) system scans application dependencies and alerts on known security vulnerabilities. Exosphere uses GitLab CI\cite{noauthor_gitlabci_nodate} with two scanners, namely Gemnasium and retire.js. Due to a technical limitation with GitLab, vulnerability reports are currently only available to project maintainers, but we plan to work around this and make the reports public.
        \end{itemize}
\end{itemize}

% - Stability
%   - Reliable instance provisioning minimizes user frustration and operator support burden
%   - No runtime exceptions in client
%   - No custom back end (for core functionality) where things break between client and infrastructure
% - Velocity/pleasure of development
%   - Extremely short tweak-test cycle for entire app (a couple of seconds)
%   - Continuous deployment
%     - Allows rapid iteration on user feedback, quick bug fixes, and cheap experimentation
%   - Browser developer tools expose all API calls to OpenStack
%   - Only one (small) programming language to learn to become productive
%     - Same language for both user interface and application logic
%   - Compiler eliminates several classes of bugs
%     - Delivers level of assurance that would require extremely diligent code testing effort in most other languages
%     - Allows fearless refactoring (and accepting of community contributions)

\subsection{Non-features}

We do not plan to implement the following, as they are incompatible with the project's goals.

\begin{itemize}
\item Requiring custom (back-end) services to support core features, to the extent that we can avoid it. (See Challenges section, subsection "Browser-accepted TLS connections to cloud instances".)
\item Re-inventing that which is both time-consuming and not uniquely valuable, e.g.
    \begin{itemize}
        \item User authentication (OpenStack handles this for us, but see "Institutional Single Sign-on" below)
        \item User/resource pools (currently using OpenStack projects)
    \end{itemize}
\end{itemize}

\subsection{Planned features}

\begin{itemize}

\item Planned features for users:

\begin{itemize}
\item Allow user to deploy a graphical desktop environment to instances launched from popular Linux distributions; provide remote graphical session using Apache Guacamole\cite{noauthor_apache_nodate}.
\item For clouds that offer GPU-connected instances (such as Jetstream 2), we also plan to support 3D-accelerated streaming desktop environments, using a VirtualGL and TurboVNC back-end for Guacamole. Strudel\cite{noauthor_strudel_nodate} is a client application which provides GPU-accelerated streaming desktops on certain HPC systems, but we are unaware of an existing solution for research cloud infrastructure, where the user has the flexibility of full root access to a Linux virtual machine. Exosphere will meet that need.
\item Ease the tasks of data management, with a web-based interface to upload and download files to/from cloud instances. This feature will likely employ the graphical file browser included with Guacamole.
\item Docker/Singularity container integration (see Challenges section below).
\item One-click support for deploying common data analysis workbenches (such as JupyterLab and RStudio) when creating an instance. Exosphere will likely pass cloud-init userdata to the instance which will install the software, and use a proxy server to deliver secure remote access to the workbench in the user's web browser.
\item Smoother installation/upgrade process for the desktop version of Exosphere, with signed binaries and fully automatic updates.
\item Custom workflow/toolchain/stack sharing, which could include, but is not limited to custom disk images.
\end{itemize}

\item Planned features for infrastructure operators:

\begin{itemize}
    \item Institutional single sign-on with OpenStack credential management/leases.
    \item Allocation service and tools.
    \item Reporting tools for operators and principal investigators of projects.
    \item Integration with user support and ticketing systems (e.g. Intercom, Zendesk, Jira).
    \item On-premises reverse proxy server (when OpenStack services are behind a firewall).
\end{itemize}

\end{itemize}

\section{Science Use Cases}

Exosphere is used by a variety of scientific and engineering contexts to provide both production and development computing environments and support reproducible research. Here we briefly describe a few.

Research groups use Exosphere to deploy and provision virtual machines for development and testing of reproducible analyses\cite{lebauer2020diag}. This core use case allows all researchers to share common computing environments as well as scale up analyses using available cloud resources.   

The PEcAn Project framework for ecological model-data synthesis\cite{lebauer2013facilitating}. PEcAn includes simulation models, statistical analyses, a database, file management, web interfaces and APIs to support the use and analysis of crop and ecosystem models. Individual components can be orchestrated using docker-compose. This is deployed on a virtual server created by Exosphere on CyVerse's OpenStack.

Exosphere is also used to provision databases that index metadata collected from small Unoccupied Aerial Systems (sUAS, i.e. drones). The ElasticSearch stack (including Kibana) is deployed on virtual servers created by Exosphere. These systems are also used to provide a variety of applications to researchers using sUAS, including plant sciences, agricultural research, and earth science (Geomorphology).

Exosphere is also used to enable educational workshops hosted by Jetstream Cloud\cite{noauthor_jetstream_reu_nodate} under the NSF REU program.

\section{Challenges}

\subsection{Same-origin policy}
Exosphere communicates with OpenStack REST/HTTP APIs on behalf of the user, but modern browsers impose the same-origin policy\cite{noauthor_same-origin_nodate}, a security feature which prevents Exosphere from making these API calls unless the OpenStack administrator configures cross-origin resource sharing\cite{noauthor_cross-origin_nodate}. This threatens users' ability to access any OpenStack deployment with Exosphere running in a web browser; the administrator of that OpenStack would need to allow the cross-origin requests.

% cmart response to Jeremy feedback: I added a paragraph break and turned this into an unordered list. I think bullets make sense here.

We found two workarounds for this issue, to avoid the need for custom configuration of OpenStack:

\begin{itemize}
    \item Exosphere can be served alongside a reverse proxy server\cite{noauthor_docscors-proxymd_nodate} which makes OpenStack API calls on behalf of the client. This introduces a back-end dependency which is lightweight but still suboptimal, given our goal to distribute and consume Exosphere as self-contained, standalone client.
    \item When Exosphere is packaged as a desktop client using Electron (rather than served to a web browser), there is no same-origin policy and no restriction of where the client makes API calls.
\end{itemize}

\subsection{Browser-accepted TLS connections to cloud instances}

In order to deliver easy, one-click interactivity to a user's cloud instances in a secure manner, we must serve content from users' instances using a TLS certificate\cite{noauthor_public_2020} that the user's web browser will accept. This is needed for a command-line shell and graphical desktop session to each instance; it is also necessary for secure connections to browser-based interactive tools like JupyterLab.

A service like Let's Encrypt\cite{noauthor_lets_nodate} could support automatic deployment of certificates to users' instances, but our attempts have not succeeded for two reasons.

First, Let's Encrypt requires the ACME\cite{kasten_automatic_nodate} client (i.e. the Exosphere user's cloud instance) to demonstrate control of a public DNS hostname. A DNS record for the instance's public IP address can be pre-created by the OpenStack administrator (perhaps en masse for the entire public IP space of that cloud), but this frustrates our goal of no administrator configuration work required in order for Exosphere to be usable with a given OpenStack cloud. We could require the Exosphere user to register a domain and create a DNS record (pointing it to the cloud instance), but that would not be user-friendly.

Second, Let's Encrypt enforces rate limits\cite{noauthor_rate_nodate} which restrict the number of certificates obtained per registered domain (currently 50 per week). If this limit were not imposed, a large OpenStack provider could pre-create public hostnames for its entire public IP space in the form of (e.g.) "instance-128-196-65-75.example-openstack-cloud.org", and a user's cloud instances could automatically obtain TLS certificates for these hostnames at first boot. Under this limit, however, a high-traffic OpenStack provider would quickly exhaust the 50 certificates that Let's Encrypt will issue in a given week for host records belonging to example-openstack-cloud.org. Unless the cloud administrator was willing to register many domains (perhaps one for each public IP address, which would be costly), this arrangement cannot scale.

\begin{figure}[tbh!]
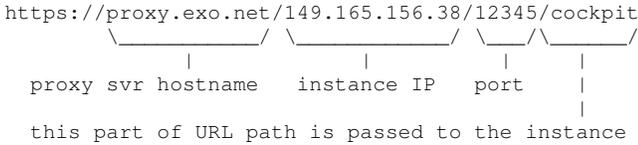

\begin{center}
\begin{lstlisting}[basicstyle=\ttfamily\footnotesize]
https://proxy.exo.net/149.165.156.38/12345/cockpit
        \___________/ \____________/ \___/\______/
              |             |          |     |
  proxy svr hostname   instance IP   port    |
                                             |
  this part of URL path is passed to the instance
\end{lstlisting}
\end{center}
\caption{URL scheme for cloud-specific TLS proxy server.}
\label{figproxy}
\end{figure}

A different approach uses a cloud-specific reverse proxy server, for content served by all Exosphere-launched instances in that cloud. The proxy server only needs one TLS certificate, and it can re-terminate TLS for all instances hosted on the same OpenStack cloud (Fig.~\ref{figproxy}), so long as the cloud administrator can assure another malicious user cannot execute a man-in-the-middle attack\cite{noauthor_man---middle_2020} on the upstream traffic between the reverse proxy server and the Exosphere user's instance.

This has been implemented\cite{noauthor_cloud-specific_nodate} but is not yet in production. The team is also exploring other approaches which do not require a TLS-terminating proxy server at each OpenStack cloud.

\subsection{Scope and boundaries of Exosphere's focus}

It is important to communicate what Exosphere does and does not do: Exosphere delivers a user-friendly way to create, manage, and access virtual computers (, storage, etc.) on OpenStack clouds, but Exosphere is not currently involved at the level of applications and analyses that users run on these cloud resources, especially not after those resources are created. This delineation of Exosphere's scope will help guide domain scientists (who are typically not IT infrastructure engineers) toward assembling a suite of tools and platforms that achieve their goals.

There are many emerging ways to build, package, and distribute scientific computing workloads: Docker, JupyterLab, and Makeflow, to name a few. As a client for cloud computing, Exosphere can support researchers' use of any of these tools on any cloud infrastructure. It is unresolved which (if any) of these tools Exosphere should integrate more tightly with, and what a useful integration would look like. We want to make Docker/Singularity containers "first-class citizens"\cite{noauthor_minimal_nodate} in Exosphere, but we have yet to determine the best way to offer this to users.

\subsection{Support for commercial cloud platforms}

OpenStack is the dominant open-source, self-hostable cloud operating system used by many research-focused clouds\cite{noauthor_openstack_2018}, but proprietary cloud platforms (e.g. Amazon Web Services, Google Cloud Platform) now dominate the commercial market for infrastructure-as-a-service. Many researchers use these commercial platforms, and some receive discounted credits. Collaborators at University of Arizona, Indiana University, and others in the community have expressed strong interest for Exosphere to support these platforms. Exosphere could help researchers exploit instance spot pricing and avoid paying for resources when they are no longer needed.

To build this support would be consistent with the goals of Exosphere, though it represents a significant increase in complexity of the application, and it may dilute the developers' focus on providing an excellent user experience for OpenStack. Each commercial cloud platform has its own proprietary API that Exosphere would need to support, though there exist projects like Apache Libcloud\cite{foundation_apache_nodate} and pkgcloud\cite{noauthor_pkgcloud_nodate} which try to abstract over these various APIs, and that may help the effort. Commercial cloud services like DigitalOcean\cite{noauthor_digitalocean_nodate} and Amazon Lightsail\cite{noauthor_amazon_nodate} already focus on user-friendliness in the way that Exosphere does, but there is no equivalent for the OpenStack ecosystem aside from Exosphere.

\subsection{Support for security-focused workloads}

Community members have expressed\cite{noauthor_security_nodate} a desire for security enhancements to support workloads on sensitive and regulated data. The Exosphere developers intend for all features to be secure by default, but some features (namely the Cockpit-powered terminal and server dashboard) currently require users' instances to be reachable via a public IP address, and that connectivity is often blocked for sensitive/secure workloads. Exosphere could be adapted to, e.g., connect to instances through a hardened bastion host\cite{noauthor_bastion_2020}, or avoid setting a local user password for recovery purposes.

Different communities will have different security requirements that may, at times, conflict with each other; a modular and configurable approach is probably necessary in order to satisfy them.

\subsection{Nomenclature}
Researchers from diverse (and often non-computational) backgrounds use different language to describe the same concepts. Exosphere should use language that users are most likely to understand, but it's often unclear which terms are most likely to be understood. Does a user create a "server", an "instance", or a "virtual computer"? Does the OpenStack concept of "project" match a user's idea of a "project"? This issue is tracked\cite{noauthor_cloud_nodate} on the Exosphere GitLab project. Likely, the team will choose the terms that are most likely to be understood by a wide audience, make them consistent within the app, and provide a good glossary.

\section{Benchmarks}
We did a preliminary benchmarking study (Table 1) which compared Exosphere to Atmosphere and Horizon. We tested the following:

\begin{itemize}
\item The time taken to deploy an instance.
\item The number of steps/clicks required to create an instance.
\end{itemize}

We looked at the time taken from start to finish for the instance to become accessible. Qualitatively, Exosphere and Horizon took similar times to provision a usable instance, depending how the user wishes to access their instance. The user could log into their Exosphere-launched instance via SSH (to the floating IP address) at 1 minute or log into their web-based terminal at 5 minutes. The instance that was launched via Horizon was available between those times, at 3 minutes. Atmosphere took 5 times as long (15 minutes) to complete deployment.

When comparing the number of steps/clicks to provision an instance on the three platforms, Exosphere and Atmosphere require fewer clicks compared to Horizon. This is in the best case where the router, network, subnet, and security groups are already set up on Horizon. A new user would have to set up these items before creating an instance, increasing the total number of clicks and deploy time. 

Overall, this is in line with Exosphere's goal of combining the ergonomics of Atmosphere with the deployment speed of Horizon. 

We will be collecting more samples using browser automation and adding the methods and results to a public repository\cite{sudarshan_cloud_2020,sanjana_sudarshan_cloud-ui-benchmarks_nodate}.

\begin{table}[ht]
\caption{Preliminary Benchmarking.}
\label{tab:label}
\setlength{\tabcolsep}{4pt}
\begin{tabular}{ r c c c }
\hline
& Atmosphere & Exosphere & Horizon \\
\hline
 Instance created (minutes) & 1.6 & 0.8 & 2\\
 Floating IP assigned (minutes)	& 2.5 & 1 & 3\\
 Instance responds to ping (minutes) & 2.5 & 1 & 3\\
 SSH/shell/desktop available (minutes)  & 15 & 5 & 3\\
 Number of clicks to create instance & 3 & 3 & 10\\
\hline
\end{tabular}
\end{table}

\section{Acknowledgements}

\begin{itemize}
    \item Jetstream Cloud, for allocation of compute resources that we're using to develop Exosphere
    \item CyVerse, for allocation of compute resources, for prior art and inspiration, and for hosting \url{https://try.exosphere.app}
    \item Andrew Lenards and Vasanth Pappu, for valuable code contributions
    \item Community contributors who provided user interviews\cite{noauthor_user_nodate}
    \item Jeremy Fischer at Indiana University for helpful review and feedback on this publication
\end{itemize}

% References
\bibliography{conference} % bibliography data in conference.bib
\bibliographystyle{IEEEtran} % makes bibtex use IEEEtran.cls

\end{document}